\documentclass[aoas,nameyear,dvips]{arximspdf}
\usepackage{graphics}

\doi{10.1214/08-AOAS220}
\volume{3}
\issue{4}
\pubyear{2009}
\firstpage{1616}
\lastpage{1633}

\makeatletter
\renewcommand{\citep}[1]{[\citeauthor{#1} (\citeyear{#1})]}
\newcommand{\h}{\hspace*{4.4pt}}

\newtheorem{lemma}{Lemma}

\newproclaim{definition}{Definition}
\newproclaim{remarks}{Remarks}

\def\sfrac#1#2{#1/#2}

\newcommand{\eqref}[1]{(\ref{#1})}

\makeatother

\begin{document}
\begin{frontmatter}

\title{Approximate null distribution of the largest root in multivariate analysis\protect\thanksref{TZ}}
\runtitle{Largest characteristic root}
\thankstext{TZ}{Supported in part by NSF DMS-05-05303, 09-06812 and NIH RO1 EB 001988.}
\begin{aug}
\author[A]{\fnms{Iain M.} \snm{Johnstone}\corref{}\ead[label=e1]{imj@stanford.edu}}
\runauthor{I. M. Johnstone}
\affiliation{Stanford University}
\address[A]{Department of Statistics\\
Stanford University\\
Stanford, California 94305\\
USA\\
\printead{e1}} 
\end{aug}

\received{\smonth{10} \syear{2008}}

\begin{abstract}
The greatest root distribution occurs everywhere in classical
multivariate analysis, but even under the null hypothesis the exact
distribution has required extensive tables or special purpose
software.
We describe a simple approximation, based on the Tracy--Widom
distribution, that in many cases can be used instead of tables or
software, at least for initial screening. The quality of approximation
is studied, and its use illustrated in a variety of setttings.
\end{abstract}


\end{frontmatter}

\setcounter{footnote}{1}
\section{Introduction}\label{sec:intro}

The greatest root distribution is found everywhere in classical
multivariate
analysis. It describes the null hypothesis distribution for the union
intersection test
for any number of classical problems, including multiple response
linear regression, MANOVA, canonical correlations, equality of
covariance matrices and so on.  However, the exact null distribution
is difficult to calculate and work with, and so the use of extensive
tables or special purpose software has always been necessary.

This paper describes a simple asymptotic approximation, based on the
Tracy Widom distribution.  The approximation is not solely asymptotic;
we argue that it is reasonably accurate over the entire range of the
parameters. ``Reasonably accurate'' means, for example, less than ten
percent relative error in the 95th percentile, even when working with
\textit{two} variables and any combination of error and hypothesis
degrees of freedom.

This paper focuses on the approximation, its accuracy
and its applicability to a range of problems in multivariate
analysis.
A companion paper \citep{john08} contains all proofs and additional discussion.

Our main claim is that for
many applied purposes, the Tracy--Widom approximation
can often, if not quite always, substitute for the elaborate tables
and computational procedures that have until now been needed.
Our hope is that this paper might facilitate the use of the
approximation in applications in conjunction with appropriate
software.

\subsection{A textbook example}\label{sec:texteg}

To briefly illustrate the Tracy--Widom approximation in action, we
revisit the rootstock data, as discussed in \citeauthor{renc02} (\citeyear{renc02}), pages~170--173.
In a classical experiment carried out from 1918--1934, apple trees of
different rootstocks were compared (\citeauthor{anhe85} [(\citeyear{anhe85}), pages~357--360] has
more detail). \citet{renc02} gives data for eight trees from each of
six rootstocks. Four variables are measured for each tree:
\texttt{Girth4}${}={}$trunk girth at 4 years in mm, \texttt{Growth4}${}={}$extension growth at 4 years in m, \texttt{Girth15}${}={}$trunk girth at 15
years in mm, and \texttt{Wt15}${}={}$weight of tree above ground at 15
years in lb.

\begin{table}[h]
\texttt{
\begin{tabular}{lccccc}
& Stock & Girth4 &  Growth4 & Girth15 &  Wt15  \\
\phantom{0}1  &    \phantom{V}I  &  111   & 2.569    & 358     & 760    \\
\phantom{0}2  &    \phantom{V}I  &  119   & 2.928    & 375     & 821    \\
$\cdots$ & &        &          &         &        \\
47 &   VI  &  113   & 3.064    & 363     & 707    \\
48 &   VI  &  111   & 2.469    & 395     & 952    \\
\end{tabular}
}
\end{table}

A one-way multivariate analysis of variance can be used to examine the
hypothesis of equality of the four-dimensional vectors of mean values
corresponding to each of the six groups (rootstocks).  The standard
tests are based on the  eigenvalues of $\mathbf{(W+B)}^{-1}
\mathbf{B}$, where $\mathbf{W}$ and $\mathbf{B}$ are the sums of
squares and products matrices within and between groups respectively.
We focus here on the largest eigenvalue, with observed value
$\theta^{\mathrm{obs}} = 0.652$.  Critical values of the null distribution
depend on parameters, here $\mathsf{s}=4, \mathsf{m} = 0, \mathsf{n} =
18.5$ [using (\ref{tablepars}) below, along with the conventions of Section~\ref{sec:equalmeans} and Definition~\ref{deftheta}].
Traditionally these are found by reference to tables or
charts. Here, the 0.05 critical value is found---after manual
interpolation in those tables---to be $\theta_{0.05} = 0.377$. The approximation
(\ref{eq:thetatwapp}) of
this paper yields the approximate 0.05 critical value
$\theta_{0.05}^{\mathrm{TW}} = 0.384$, which clearly serves just as well for
rejection of the null hypothesis.

It is more difficult in standard packages to obtain $p$-values
corresponding to $\theta^{\mathrm{obs}}$. The default is to use a lower bound
based on the $F$ distribution [see (\ref{eq:p-F})], here $p_F(\theta^{\mathrm{obs}}) =
1.7 \times 10^{-8}$, which is anti-conservative and
several orders of magnitude below the Tracy--Widom
approximation given in this paper at (\ref{eq:p-TW}), $p_{\mathrm{TW}}(\theta^{\mathrm{obs}}) = 5.6 \times
10^{-5}$. The latter is much closer to the formally correct
value,\footnote{This (actually approximate) value is obtained by
interpolation from Koev's  function \mbox{\texttt{pmaxeigjacobi}}
which
only handles integer values of  $\mathsf{n}$.}
$p(\theta^{\mathrm{obs}}) = 3.7 \times 10^{-6}$. When $p$-values are very
small, typically only the order of magnitude is of interest. We suggest
in Section~\ref{secpvalues} that the Tracy--Widom approximation
generally comes close
to the correct order of magnitude, whereas the default $F$ bound is
often off by several orders.

\subsection{Organization of paper}\label{sec:organization-paper}

The rest of this introduction provides enough background to state the
main Tracy--Widom approximation result.
Section~\ref{sec:test-of-fit} focuses on the quality of the
approximation by looking both at conventional percentiles and at very
small $p$-values.
The remaining Sections~\ref{sec:test-indep-two}--\ref{sec:mlm}
describe some of the classical uses of the largest root test in
multivariate analysis, in each case in enough detail to identify the
parameters used. Some extra attention is paid in Section~\ref{sec:mlm}
to the multivariate linear model, in view of the wide variety of null
hypotheses that can be considered.

\subsection{Background}\label{sec:background}

Our setting is the distribution theory associated with sample draws
from the multivariate normal distribution.
For definiteness, we use the notation of \citet{mkb79}, to which
we also refer for much standard background material.
Thus, if $\mathbf{x}_1, \ldots, \mathbf{x}_n$ denotes a random sample
from $N_p(\bolds{\mu}, \bolds{\Sigma})$, a $p$-variate Gaussian
distribution with mean $\bolds{\mu}$ and covariance matrix
$\bolds{\Sigma}$, then we call the $n \times p$ matrix $\mathbf{X} =
(\mathbf{x}_1, \ldots, \mathbf{x}_n)'$, whose $i$th row contains
the $i$th sample $p$-vector, a normal data matrix.

A $p \times p$ matrix $\mathbf{A}$ that can be written $\mathbf{A} =
\mathbf{X}' \mathbf{X}$ in terms of such a normal data matrix is said
to have a Wishart distribution with scale matrix $\bolds{\Sigma}$ and
degrees of freedom parameter $n$,  $\mathbf{A} \sim
W_p(\bolds{\Sigma},n)$. When $p = 1$, this
reduces to a scaled chi-squared law $\sigma^2
\chi^2_{(n)}$.

We consider analogs of the $F$ and Beta
distributions of multivariate analysis, which are based on two
independent chi-squared variates.
Thus, let $\mathbf{A} \sim W_p(\bolds{\Sigma},m)$ be independent of
$\mathbf{B} \sim W_p(\bolds{\Sigma},n)$.
If  $m \geq p$, then $\mathbf{A}^{-1}$ exists and the
nonzero eigenvalues of $\mathbf{A}^{-1} \mathbf{B}$ are quantities of
interest that generalize the univariate $F$ ratio.
We remark that the scale matrix $\bolds{\Sigma}$ has no effect on the
distribution of these eigenvalues, and so, without loss of generality,
we can suppose that $\bolds{\Sigma} = \mathbf{I}$.

The matrix analog of a Beta variate is based on the eigenvalues of
$(\mathbf{A}+ \mathbf{B})^{-1} \mathbf{B}$, and leads to the following:
\renewcommand{\thedefinition}{$\theta$}
\begin{definition}[{[\citeauthor{mkb79} (\citeyear{mkb79}), page~84]}]\label{deftheta}
Let $\mathbf{A} \sim\break W_p(\mathbf{I},m)$ be independent of
$\mathbf{B} \sim W_p(\mathbf{I},n)$, where $m \geq p$.
Then the largest eigenvalue $\theta$ of $(\mathbf{A}+ \mathbf{B})^{-1}
\mathbf{B}$ is called the \textit{greatest root statistic} and its
distribution is denoted $\theta(p,m,n)$.
\end{definition}

Since $\mathbf{A}$ is positive definite, we have $0 < \theta < 1$.
Clearly $\theta(p,m,n)$ can also be defined as the largest root of the
determinantal equation
\[
\det [ \mathbf{B} - \theta (\mathbf{A} + \mathbf{B})] = 0.
\]

Specific examples will be given below, but in general the parameter
$p$ refers to dimension, $m$ to the ``error'' degrees of freedom and
$n$ to the ``hypothesis'' degrees of freedom. Thus, $m+n$ represents
the ``total'' degrees of freedom.

There are $\min(n,p)$ nonzero eigenvalues of $\mathbf{A}^{-1}
\mathbf{B}$ or, equivalently, $\min(n,p)$ nonzero roots $\theta =
(\theta_i)$ of the determinantal equation above. The joint density
function of these roots is given by
\begin{equation}\label{eq:jtden}
p(\theta) = C \prod_{i=1}^{\min(n,p)} \theta_i^{(|n-p|-1)/2}
(1-\theta_i)^{(m-p-1)/2} \Delta(\theta),
\end{equation}
where $\Delta(\theta) = \prod_{i \neq j} |\theta_i - \theta_j|$ (see,
e.g., \citeauthor{muir82} [(\citeyear{muir82}), page~112], or \citeauthor{ande03} [(\citeyear{ande03}), pages~536--537]).
We shall not need the explicit form of the density in this paper; it
is, however, useful sometimes in matching up the various parameter
choices used in different references and packages.

The greatest root distribution has the property
\[
\theta(p,m,n) \stackrel{\mathcal{D}}{=} \theta(n, m+n-p, p),
\]
useful, in particular, in the case when $n < p$
[e.g. \citeauthor{mkb79} (\citeyear{mkb79}), page~84].

\subsection{Main result}\label{sec:main-result}

Empirical and theoretical investigation has
shown that it is useful to develop the approximation in terms of the
logit transform of $\theta$; thus, we define
\begin{equation}\label{eq:logit}
W(p,m,n) = \operatorname{logit} \theta(p,m,n)
= \log  \biggl( \frac{ \theta(p,m,n)}{1 - \theta(p,m,n)}  \biggr).
\end{equation}

Our main result, stated more formally below, is that with appropriate
centering and scaling, $W$ is approximately Tracy--Widom distributed:
\begin{equation}\label{eq:TWlim}
\frac{W(p,m,n) - \mu(p,m,n)}{\sigma(p,m,n)}
  \stackrel{\mathcal{D}}{\Rightarrow}   F_1.
\end{equation}
The centering and scaling parameters are defined by
\begin{eqnarray}\label{eq:mu}
\mu(p,m,n) & = & 2 \log \tan  \biggl( \frac{\phi+\gamma}{2}  \biggr),\\\label{eq:sigma}
\sigma^3(p,m,n)
& = &\frac{16}{(m+n-1)^2}   \frac{1}{\sin^2(\phi+\gamma) \sin\phi \sin \gamma},
\end{eqnarray}
where the angle parameters $\gamma, \phi$ are defined by
\begin{eqnarray*}
\sin^2  \biggl( \frac{\gamma}{2}  \biggr)
& = & \frac{ \min(p,n) - \sfrac{1}{2}}{m+n-1},  \\
\sin^2  \biggl( \frac{\phi}{2}  \biggr)
& = & \frac{ \max(p,n) - \sfrac{1}{2}}{m+n-1}.
\end{eqnarray*}

\subsection{More on the Tracy--Widom law}\label{sec:more-tracy-widom}

\begin{figure}

\includegraphics{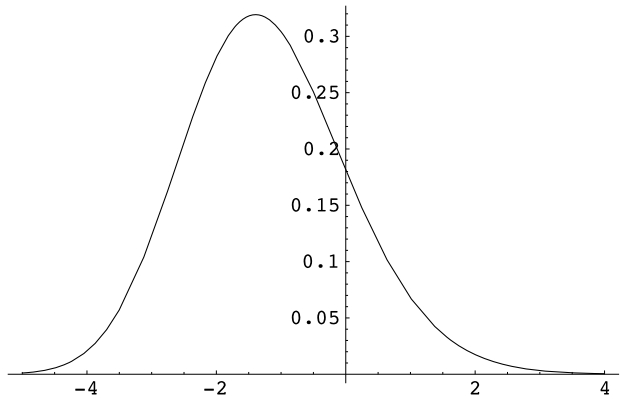}

\caption{Density of the Tracy--Widom distribution $F_1$.}\label{fig:tw1}
\end{figure}

The $F_1$ distribution, due to\break \citet{trwi96} and
plotted in Figure~\ref{fig:tw1}, has its
origins in mathematical physics---see \citet{trwi96}; \citet{john00c} for
further details.
The density is asymmetric, with mean $\doteq -1.21$ and SD $\doteq
1.27$.
Both tails have exponential decay, the left tail like $e^{-|s|^3/24}$
and the right tail like $e^{-(2/3) s^{3/2}}$.

For the present paper, what is important is that the $F_1$
distribution does not depend on any parameters, and the distribution
itself, along with its inverse and percentiles, can be tabulated as
univariate special functions. These functions play the same role in
this paper as the standard normal distribution $\Phi$, its inverse
$\Phi^{-1}$ and percentiles $z_\alpha$ play in traditional statistical
application.

\subsubsection*{Software}
An \texttt{R} package \texttt{RMTstat} is available at CRAN
(\href{http://cran.r-project.org}{cran.r-project.org}).
It facilitates computation of the distributional approximations and largest
root tests described in this paper,
and the use of percentiles and random draws from the $F_1$
distribution.
Its scope and use is described in more detail in an
accompanying report \citet{jmps10}.
A parallel \texttt{MATLAB} package is in development; it will also
contain code to reproduce the figures and table in this paper.


\subsubsection*{Percentiles}
Let $f_\alpha$ denote the $\alpha$th
percentile of $F_1$. For example,
\[
f_{0.90} = 0.4501,\qquad
f_{0.95} = 0.9793,\qquad
f_{0.99}  = 2.0234.
\]
Then the $\alpha$th percentile of $\theta(p,m,n)$ is given
approximately by
\begin{equation}\label{eq:thetatwapp}
\theta_\alpha = e^{\mu + f_\alpha \sigma}/(1 + e^{\mu + f_\alpha \sigma}),
\end{equation}
where $\mu = \mu(p,m,n), \sigma = \sigma(p,m,n)$ are given by
(\ref{eq:mu}) and (\ref{eq:sigma}).

The more formal statement of (\ref{eq:TWlim}) goes as follows.
Assume $p, m$ and $n \rightarrow \infty$ together in such a way that
\begin{equation}\label{eq:limits}
\lim \frac{ p \wedge n}{m+n} > 0,\qquad
\lim \frac{m}{p} > 1.
\end{equation}
For each $s_0 \in \mathbb{R} $,
there exist $c, C > 0$
such that for $s \geq s_0$,
\[
| P \{ W(p,m,n) \leq \mu(p,m,n) + \sigma(p,m,n) s \} - F_1(s) | \leq C
p^{-2/3} e^{-cs}.
\]
For the full proof and much more discussion and detail, see the companion
paper \citep{john08}.

\begin{remarks*}
\textit{Smallest eigenvalue.}
If $\mathbf{A}$ and $\mathbf{B}$ are as in the definition of
$\theta(p,m,n)$, then let $\tilde \theta(p,m,n)$ denote the \textit{smallest}
eigenvalue of $(\mathbf{A+B})^{-1} \mathbf{B}$. Its distribution is
given by
\[
\tilde \theta(p,m,n) \stackrel{\mathcal{D}}{=} 1 - \theta(p,n,m),
\]
(note the reversal of $m$ and $n$!).
In particular, the Tracy--Widom distribution will give a generally
useful approximation to the \textit{lower} tail of $\tilde
\theta(p,m,n)$.

\textit{Complex-valued data}.
There is an entirely analagous result when $\mathbf{A}$ and
$\mathbf{B}$ follow complex Wishart distributions, with a modified
limit distribution $F_2$. Details are given in \citet{john08}.
\end{remarks*}


\section{Quality of approximation}\label{sec:test-of-fit}

\subsection{Comparison with percentiles}\label{sec:percentiles}

There is a substantial literature computing percentage points of the
greatest root distribution for selected parameter values, partially
reviewed below.
The standard paramaterization used in these tables arises from writing
the joint density of the roots $\theta_i$ as
\[
p(\theta) = C \prod_{i=1}^\mathsf{s} \theta_i^\mathsf{m}
(1-\theta_i)^\mathsf{n} \Delta(\theta).
\]
From this and (\ref{eq:jtden}) it is apparent that our ``MKB''
parameters $(p,m,n)$ are related to the ``Table'' parameters
$(\mathsf{s}, \mathsf{m}, \mathsf{n})$ via
\begin{eqnarray}
\mathsf{s} & = & \min(n,p),\qquad\hspace*{27.5pt}         p  = \mathsf{s}, \nonumber \\
\mathsf{m} & = & ( |n-p|-1)/2,\qquad      m  = \mathsf{s} + 2 \mathsf{n} +1, \label{tablepars} \\
\mathsf{n} & = & ( m-p-1)/2,\qquad\hspace*{6.5pt}        n  = \mathsf{s} + 2 \mathsf{m} + 1. \nonumber
\end{eqnarray}

In terms of the table parameters and $\mathsf{N} = 2( \mathsf{s} +
\mathsf{m} + \mathsf{n}) + 1$, the centering and scaling constants of
the Tracy--Widom approximation are given by
\[
\sin^2  \biggl( \frac{\gamma}{2}  \biggr)
= \biggl( \mathsf{s} - \frac{1}{2}\biggr)\big/ \mathsf{N},\qquad
\sin^2  \biggl( \frac{\phi}{2}  \biggr)
= \biggl( \mathsf{s} + 2 \mathsf{m} + \frac{1}{2}\biggr)\big/ \mathsf{N}
\]
and
\begin{equation}\label{eq:musigSAS}
\mu  = 2 \log \tan  \biggl( \frac{\phi+\gamma}{2}  \biggr),\qquad
\sigma^3  =
\frac{16}{\mathsf{N}^2}   \frac{1}{\sin^2(\phi+\gamma) \sin
\phi \sin \gamma}.
\end{equation}

We turn to the comparison of percentage points $\theta_\alpha^{\mathrm{TW}}$
from the Tracy--Widom approximation (\ref{eq:thetatwapp}) with the
exact values $\theta_\alpha$ for small values of the table parameters
$(\mathsf{s}, \mathsf{m}, \mathsf{n})$. The most extensive tabulations
of $\theta_\alpha(\mathsf{s}, \mathsf{m}, \mathsf{n})$ have been made
by William Chen; he has graciously provided the author with the
complete version of the tables excerpted in
\citeauthor{chen02a} (\citeyear{chen02a,chen03,chen04,chen04a}).

\begin{figure}

\includegraphics{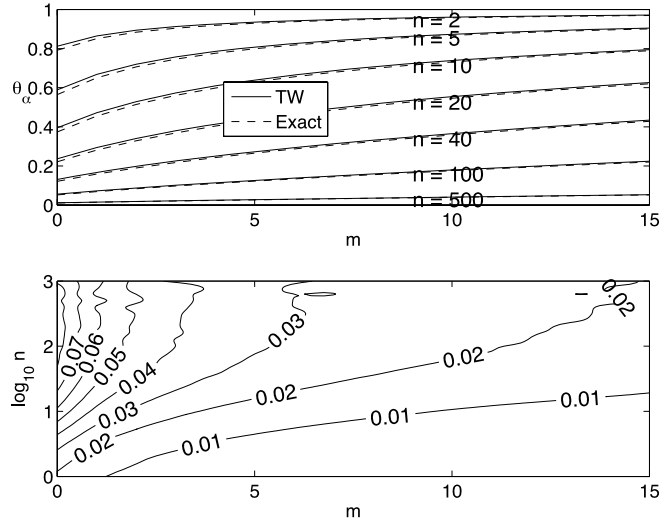}

\caption{Comparison of exact and approximate 95th
percentiles for $\mathsf{s}=2$. Top panel: solid line is
the Tracy--Widom approximation $\theta_\alpha^{\mathrm{TW}}(2,\mathsf{m},
\mathsf{n})$ plotted as a function of $\mathsf{m}$ for values of
$\mathsf{n}$ shown. Dashed lines are the exact percentiles
$\theta_\alpha(2,\mathsf{m},\mathsf{n})$ from Chen's tables.
Bottom panel: Contour plots of relative error $r =
(\theta_\alpha^{\mathrm{TW}}/\theta_\alpha) - 1$. Horizontal axis is
$\mathsf{m}$, vertical axis is $\log_{10} \mathsf{n}$, thus
covering the range from $\mathsf{n}=1$ to $1000$.}\label{fig:twcompare1}
\end{figure}

\begin{figure}

\includegraphics{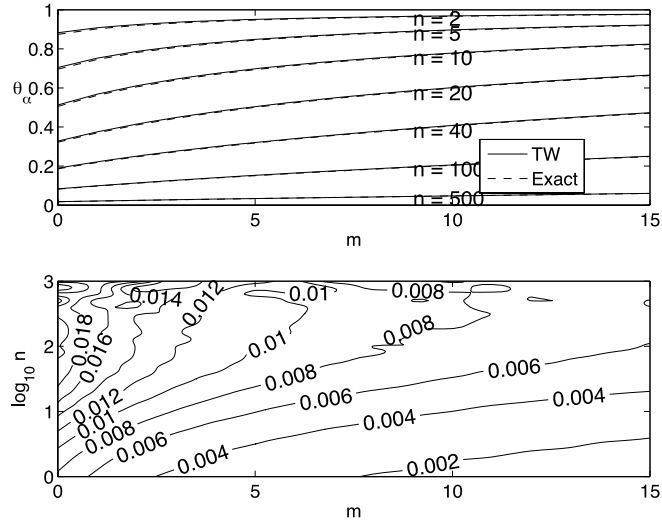}

\caption{Comparison of exact and approximate 90th
percentiles for $\mathsf{s}=4$. Top panel: solid line is
the Tracy--Widom approximation $\theta_\alpha^{\mathrm{TW}}(4,\mathsf{m},
\mathsf{n})$ plotted as a function of $\mathsf{m}$ for values of
$\mathsf{n}$ shown. Dashed lines are the exact percentiles
$\theta_\alpha(4,\mathsf{m},\mathsf{n})$ from Chen's tables.
Bottom panel: Contour plots of relative error $r =
(\theta_\alpha^{\mathrm{TW}}/\theta_\alpha) - 1$. Horizontal axis is
$\mathsf{m}$, vertical axis is $\log_{10} \mathsf{n}$, thus
covering the range from $\mathsf{n}=1$ to $1000$.}\label{fig:twcompare2}
\end{figure}

Figures~\ref{fig:twcompare1} and~\ref{fig:twcompare2} plot
$\theta_\alpha^{\mathrm{TW}}$ against $\theta_\alpha$ at 95th and
90th percentiles for $\mathsf{s}=2$. This is the smallest
relevant value of $\mathsf{s}$---otherwise we are in the univariate
case covered by $F$ distributions. The bottom panels, in particular,
focus on the \textit{relative error}
\[
r = (\theta_\alpha^{\mathrm{TW}}/\theta_\alpha) - 1.
\]
Figure~\ref{fig:twcompare1} shows that even for $\mathsf{s}=2$, the
95th percentile of the TW approximation has a relative error of
less than 1 in 20 except in the zone where both $\mathsf{m}\leq 2$ and
$\mathsf{n} \geq 10$, where the relative error is still less than 1 in
10.
Note that the relative error is always positive in sign, implying that
the approximate critical points yield a conservative test.
More extensive contour plots covering $\mathsf{s} = 2(1)6$ and
90th, 95th and 99th percentiles may
be found in \citet{joch07}.

%
%

\subsubsection*{Work on tables}
There has been a large amount of work to prepare tables or charts for the
null distribution of the largest root, much of which is reviewed in
\citet{chen03}.
We mention contributions by the following:
\citeauthor{nand48} (\citeyear{nand48,nand51});
\citeauthor{fore57} (\citeyear{fore57});
\citeauthor{fost57} (\citeyear{fost57,fost58});
\citeauthor{pill55} (\citeyear{pill55,pill56,pill56a,pill57,pill65,pill67});
\citeauthor{pill59} (\citeyear{pill59});
\citet{heck60};
\citet{kris80};
\citet{pifl84};
\citeauthor{chen02a} (\citeyear{chen02a,chen03,chen04,chen04a}).

Because of the dependence on the three parameters, these tables can
run up to 25 pages in typical textbooks, such as those of
\citet{jowi07} and \citet{morr05}.






%



\subsubsection*{Code}
\citet{cons63} expresses the c.d.f. of the largest root distribution in
terms of a matrix hypergeometric function.
\citet{koed06}~have developed efficient algorithms (and a MATLAB
package available at \url{http://www-math.mit.edu/\textasciitilde plamen}) for the
evaluation of such matrix hypergeometric functions using recursion
formulas from group representation theory.

\citet{koev07} collects useful formulas and explains how to use them
and \texttt{mhg} to compute the exact c.d.f. and percentiles for the
largest root distribution over a range of values of the ``MKB''
parameters corresponding to  $m,n,p \leq 17$, and
$m,n,p \leq 40$ when $n-p$ is odd.


SAS/STAT 9.0 made available an option for computing
exact $p$-values using \citet{davi72}; \citet{pifl84}. There is also some
stand-alone software described by \citeauthor{lutz92} (\citeyear{lutz92,lutz00}).

\subsection{Accuracy of $p$-values}\label{secpvalues}

\textit{The univariate} $F$ \textit{bound}.
We recall the hypothesis that $\mathbf{A} \sim W_p(\mathbf{I},m)$ be
distributed independently of $\mathbf{B} \sim W_p(\mathbf{I},n)$, and the
characterization of the largest eigenvalue given by
\begin{equation}\label{eq:maxim}
\lambda_{\max}( \mathbf{A}^{-1} \mathbf{B} )
=  \max_{|\mathbf{u}| = 1} \frac{\mathbf{u}' \mathbf{B} \mathbf{u}}{
\mathbf{u}' \mathbf{A} \mathbf{u}}.
\end{equation}
For fixed $\mathbf{u}$ of unit length, the numerator and denominator
are distributed as independent $\chi^2_{(n)}$ and $\chi^2_{(m)}$
respectively, and so, again for fixed $\mathbf{u}$, the ratio has an
$F_{n,m}$ distribution. Consequently, we have the simple bound
\[
\frac{m}{n} \lambda_{\max}( \mathbf{A}^{-1} \mathbf{B} )
> F \sim F_{n,m}.
\]

Using the $F_{n,m}$ distribution in place of the actual greatest root
law yields a lower bound for the significance level, or $p$-value.  We
shall see that this bound can be anti-conservative by several orders of
magnitude, leading to overstatements of the empirical evidence against
the null hypothesis.   And furthermore, one can expect that the higher the
dimension $p$ of the search space in (\ref{eq:maxim}), the worse the
bound provided by the $F$ distribution.

The default $p$-value provided in both SAS and \textsf{R} (through
package \textsf{car}) uses this unsatisfactory distribution
bound.

%

Table~\ref{tab:S26tails} 
attempts to capture a variety of scenarios within the computational
range of Koev's software.

Column \texttt{Exact} shows a range of significance levels $\alpha$
covering several orders of magnitude.
Column \texttt{Largest Root} shows the corresponding quantiles
$\theta_\alpha$ of the largest root distribution, for the given values
of $(\mathsf{s,m,n})$---these are computed using Koev's MATLAB
routine \texttt{qmaxeigjacobi}.
Thus, an observed value of $\theta(\mathsf{s,m,n}) = \theta_\alpha$
would correspond to an exact $p$-value $\alpha$.

The remaining columns compare the Tracy--Widom approximation and the
$F$ bound.
The $p$-value obtained from the Tracy--Widom approximation is given by
\begin{equation}\label{eq:p-TW}
P_{\mathrm{TW}}(\theta_\alpha) = 1 - F_1\bigl( \bigl(\operatorname{logit}(\theta_\alpha) - \mu\bigr)/\sigma\bigr),
\end{equation}
where $\mu$ and $\sigma$ are computed from (\ref{eq:musigSAS}).

The $F$ bound on the $p$-value is given by
\begin{equation}\label{eq:p-F}
P\bigl( \theta(\mathsf{s,m,n}) > \theta_\alpha\bigr) >
P_F(\theta_\alpha) = 1 - F_{\nu_1,\nu_2}
\bigl( \nu_2 \theta_\alpha/\bigl(\nu_1(1-\theta_\alpha)\bigr) \bigr),
\end{equation}
where $\nu_1 = \mathsf{s} + 2\mathsf{m} +1$ and $\nu_2 = \mathsf{s}+2\mathsf{n}+1$
denote the hypothesis and error degrees of freedom respectively.

The two tables consider $\mathsf{s}=2$ and $6$ variables respectively.
The values of $\mathsf{m}=-0.5$ and $5$ correspond to $\mathsf{s}$ and
$\mathsf{s}+11$ hypothesis degrees of freedom, while the values of
$\mathsf{n}=2$ and $10$ translate to $\mathsf{s}+5$ and
$\mathsf{s}+21$ error degrees of freedom respectively.

At the $10\%$ and $5\%$ levels, the Tracy--Widom approximation is
within $20\%$ of the true $p$-value at $\mathsf{s}=6$, and within
$35\%$ of truth at $\mathsf{s}=2$. The $F$-value is wrong by a factor
of four or more at $\mathsf{s}=2$, and by three orders of magnitude at
$\mathsf{s}=6$.
At smaller significance levels, the Tracy--Widom approximation
generally stays within one order of magnitude of the correct $p$-value---except
at $\mathsf{(s,m,n)}=(2,-0.5,10)$. The $F$ approximation is off
by many orders of magnitude when $\mathsf{s}=6$.

In addition, we note that the Tracy--Widom approximation is
conservative in nearly all cases, the exception being for $\theta \geq
0.985$ in the case $\mathsf{(s,m,n)}=(6,-0.5, 2)$.
In contrast, the $F$ approximation is \textit{always}
[cf. (\ref{eq:p-F})] anti-conservative, often badly so.

In applications one is often concerned only with the general order of
magnitude of the $p$-values associated with tests of the various
hypotheses that are entertained---not least because the assumptions of
the underlying model are at best approximately true. For this
purpose, then, it may be argued that the TW approximate $p$-value is
often quite adequate over the range of $(\mathsf{s,m,n})$ values.
Of course, if $(\mathsf{s,m,n})$ is not too large and greater
precision is required, then exact $p$-values can be sought, using, for example, SAS or
Koev's software.

\begin{table}
\tabcolsep=0pt
\caption{Comparison of the Tracy--Widom approximation and $F$ bound for
cases with $\mathsf{s}= 2$ and $\mathsf{s}=6$ variables}\label{tab:S26tails}
{\fontsize{6}{8}\selectfont
\texttt{
\begin{tabular*}{\textwidth}{@{\extracolsep{\fill}}clll@{\hspace*{15pt}} clll@{}}
\hline
\textbf{Largest root}
& \multicolumn{1}{c}{\textbf{Exact}}
& \multicolumn{1}{c}{\textbf{Tracy}$\bolds{-}$\textbf{Widom}}
& \multicolumn{1}{c}{\textbf{F}}
& \textbf{Largest root}
& \multicolumn{1}{c}{\textbf{Exact}}
& \multicolumn{1}{c}{\textbf{Tracy}$\bolds{-}$\textbf{Widom}}
& \multicolumn{1}{c@{}}{\textbf{F}} \\
\hline
\multicolumn{4}{c}{{\fontsize{6.7}{8.7}\selectfont$\mathsf{s}=2,  \mathsf{m}= -0.5, \mathsf{n}= 2$}}
&\multicolumn{4}{c@{}}{{\fontsize{6.7}{8.7}\selectfont$\mathsf{s}=6, \mathsf{m}= -0.5, \mathsf{n}= 2$}}\\
0.663 &     0.1    &  \h0.119         &     0.0223      &0.918 &        0.1 &    \h0.115 &  2.23e-005 \\
0.737 &     0.05   &  \h0.066         &    0.00933      &0.938 &       0.05 &    \h0.0598 &  4.99e-006 \\
0.850 &     0.01   &  \h0.0169        &    0.00131      &0.966 &       0.01 &    \h0.0116 &  1.92e-007 \\
0.881 &     0.005  &  \h0.00927       &   0.000573      &0.973 &      0.005 &    \h0.00545 &  4.96e-008 \\
0.931 &     0.001  &  \h0.00222       &  8.49e-005      &0.985 &      0.001 &    \h0.000839 &   2.3e-009 \\
0.968 &     0.0001 &  \h0.000251      &  5.65e-006      &0.993 &     0.0001 &    \h4.35e-005 &   3.1e-011 \\
0.985 &     1e-005 &  \h2.38e-005     &  3.81e-007      &0.997 &     1e-005 &    \h1.64e-006 &  4.38e-013 \\
0.993 &     1e-006 &  \h1.89e-006     &  2.58e-008      &0.999 &     1e-006 &    \multicolumn{1}{c}{NaN} &  6.33e-015
\\[5pt]
\multicolumn{4}{c}{{\fontsize{6.7}{8.7}\selectfont$\mathsf{s}=2, \mathsf{m}= -0.5, \mathsf{n}= 10$}}
&\multicolumn{4}{c}{{\fontsize{6.7}{8.7}\selectfont$\mathsf{s}=6, \mathsf{m}= -0.5, \mathsf{n}= 10$}}\\
0.268 &        0.1 &   \h0.117 &     0.0278      &   0.597 &        0.1 &     \h0.11 &   0.000206 \\
0.318 &       0.05 &   \h0.0669 &     0.0123      &   0.633 &       0.05 &    \h0.0577 &  6.49e-005 \\
0.418 &       0.01 &   \h0.0214 &    0.00199      &   0.698 &       0.01 &    \h0.0134 &  5.46e-006 \\
0.456 &      0.005 &   \h0.0137 &   0.000919      &   0.721 &      0.005 &    \h0.00722 &  1.99e-006 \\
0.533 &      0.001 &   \h0.00522 &   0.000157      &   0.766 &      0.001 &   \h0.00172 &  2.05e-007 \\
0.624 &     0.0001 &   \h0.00146 &  1.31e-005      &   0.816 &     0.0001 &   \h0.000223 &  8.97e-009 \\
0.696 &     1e-005 &   \h0.000443 &  1.11e-006      &   0.854 &     1e-005 &  \h2.86e-005 &  4.29e-010 \\
0.755 &     1e-006 &   \h0.000141 &  9.59e-008      &   0.884 &     1e-006 &  \h3.57e-006 &  2.17e-011
\\[5pt]
\multicolumn{4}{c@{}}{{\fontsize{6.7}{8.7}\selectfont$\mathsf{s}=2, \mathsf{m}= 5, \mathsf{n}= 10$}}
&\multicolumn{4}{c@{}}{{\fontsize{6.7}{8.7}\selectfont$\mathsf{s}=6, \mathsf{m}= 5, \mathsf{n}= 10$}}\\
0.592 &        0.1 &   \h0.112 &     0.0234       &0.757 &        0.1 &      \h0.108 &   0.000117 \\
0.629 &       0.05 &   \h0.0602 &     0.0103       &0.781 &       0.05 &     \h0.0557 &  3.63e-005 \\
0.697 &       0.01 &   \h0.0149 &    0.00164       &0.823 &       0.01 &     \h0.0119 &  2.99e-006 \\
0.721 &      0.005 &   \h0.00827 &   0.000758       &0.837 &      0.005 &    \h0.00606 &  1.08e-006 \\
0.767 &      0.001 &   \h0.00215 &   0.000129       &0.864 &      0.001 &    \h0.00125 &   1.1e-007 \\
0.817 &     0.0001 &   \h0.000318 &  1.07e-005       &0.894 &     0.0001 &   \h0.000125 &  4.75e-009 \\
0.855 &     1e-005 &   \h4.71e-005 &  9.04e-007       &0.917 &     1e-005 &  \h1.17e-005 &  2.25e-010 \\
0.885 &     1e-006 &   \h6.88e-006 &  7.79e-008      &0.934 &     1e-006 &   \h1.03e-006 &  1.13e-011 \\
\hline
\end{tabular*}
}
}
\end{table}

\section{Testing for independence of two sets of variables}\label{sec:test-indep-two}

Let $\mathbf{x}_1, \ldots, \mathbf{x}_n$ be a random sample from $N_p
(\bolds{\mu}, \bolds{\Sigma})$.
Partition the variables into two sets with dimensions $p_1$ and $p_2$
respectively, $p_1 + p_2 = p$.
Suppose that $\bolds{\Sigma}$ and the sample covariance matrix
$\mathbf{S}$ are partitioned correspondingly.
We consider testing the null hypothesis of independence of the two
sets of variables: $\bolds{\Sigma}_{12} = \mathbf{0}$.
The union-intersection test is based on the largest eigenvalue
$\lambda_1$ of $\mathbf{S}_{22}^{-1} \mathbf{S}_{21}
\mathbf{S}_{11}^{-1} \mathbf{S}_{12}$
(\citet{mkb79}, page~136) and under $H_0, \lambda_1$ has the
greatest root distribution $\theta(p_2, n-1-p_1, p_1)$.
\citet{mkb79} consider an example test of independence of $n=25$ head
length and breadth measurements between first sons and second sons, so
that $p_1 = p_2 = 2$.
The observed value $\lambda_1^{\mathrm{obs}} = 0.622$ exceeds the critical
value $\theta_{0.05}=0.330$ found by interpolation from the tables.
The Tracy--Widom approximation $\theta_{0.05}^{\mathrm{TW}} = 0.356$ is found
from \eqref{eq:thetatwapp} and serves equally well for rejection of
$H_0$ in this case.

\section{Canonical correlation analysis}\label{sec:canon-corr-analys}

Again we have two sets of variables, an $\mathbf{x}$-set with $q$
variables and a $\mathbf{y}$-set with $p$ variables. The goal is to
find maximally correlated linear combinations $\eta = \mathbf{a}'
\mathbf{x}$ and $\phi = \mathbf{b}' \mathbf{y}$.
We suppose that $(\mathbf{X}, \mathbf{Y})$ is a data matrix of $n$
samples (rows) on $q + p$ variables (columns) such that each row is an
independent draw from $N_{p+q}(\bolds{\mu},\bolds{\Sigma})$.
Again let $\mathbf{S}$ be the sample covariance matrix, assumed
partitioned $\mathbf{S} =
\bigl(
{\mathbf{S}_{11} \atop \mathbf{S}_{21}} \enskip
{\mathbf{S}_{12} \atop \mathbf{S}_{22}}
\bigr)
$.
The sample squared canonical correlations $(r_i^2)$ for $i = 1,
\ldots, k = \min(p,q)$ are found as the eigenvalues of $\mathbf{M}_2 =
\mathbf{S}_{22}^{-1} \mathbf{S}_{21} \mathbf{S}_{11}^{-1}
\mathbf{S}_{12}$ [\citet{mkb79}, Sections~10.2.1 and~10.2.2].
The population squared canonical correlations $\rho_i^2$ are, in turn,
the eigenvalues of
$\bolds{\Sigma}_{22}^{-1} \bolds{\Sigma}_{21} \bolds{\Sigma}_{11}^{-1}
\bolds{\Sigma}_{12}$.
In both cases, we assume that the correlations are arranged in
decreasing order.
The test of the null hypothesis of zero correlation,
$H_0\dvtx  \rho_1 = \cdots = \rho_k =0$, is based on the largest eigenvalue
$r_1^2$ of $\mathbf{M}_2$.
Under $H_0$, it is known that $r_1^2$ has the $\theta(p, n-q-1, q)$
distribution, so that the Tracy--Widom approximation can be applied.

\subsection*{Nonnull cases---a conservative test}
Often it may be apparent that the first $k$ canonical correlations are
nonzero and the main interest focuses on the significance of
$r_{k+1}^2, r_{k+2}^2$, etc.
We let $H_s$ denote the null hypothesis that $\rho_{s+1} = \cdots =
\rho_p = 0$,  and write
$\mathcal{L}(r_k|p,q,n; \bolds{\Sigma})$ for the distribution of
the $r$th c.c. under population covariance matrix $\bolds{\Sigma}$.
When the covariance matrix $\bolds{\Sigma} \in H_s$, the
$(s+1)$st canonical correlation is stochastically smaller than the
\textit{largest} canonical correlation in a related null model:
\begin{lemma}
If $\bolds{\Sigma} \in H_s$, then
\[
\mathcal{L}(r_{s+1}| p,q,n; \bolds{\Sigma})
\stackrel{st}{<}
\mathcal{L}(r_{1}| p,q-s,n; \mathbf{I}).
\]
\end{lemma}

This nonasymptotic result follows from interlacing properties of the
singular value decomposition (\hyperref[sec:concl-disc-1]{Appendix}).
Since $\mathcal{L}(r_{1}^2| p,q-s,n; \mathbf{I})$ is given by the null
distribution $\theta(p, n+s-q-1, q-s)$, we may use the latter to
provide a conservative $p$-value for testing $H_s$.
In turn, the $p$-value for $\theta(p, n+s-q-1, q-s)$ can be
numerically approximated as in \eqref{eq:thetatwapp} using the
Tracy--Widom distribution.

\subsection*{Example}
\citet{waug42} gave perhaps the first significant illustration of CCA
using data on $n=138$ samples of Canadian Hard Red Spring wheat and
the flour made from each of these samples. The aim was to seek highly
correlated indices $\mathbf{a}'\mathbf{x}$ of wheat quality and
$\mathbf{b}'\mathbf{y}$ of flour quality, since a well correlated
grading of raw (wheat) and finished (flour) products was believed to
promote fair pricing of each. In all, $q = 5$ wheat characteristics---kernel texture, test weight, damaged kernels, foreign material, crude
protein in wheat---and $p = 4$ flour characteristics---wheat per
bushel of flour, ash in flour, crude protein in flour, gluten quality
index---were measured.
The resulting squared canonical correlations were
$(r_1^2, r_2^2, r_3^2, r_4^2) = (0.923, 0.554,\break 0.056, 0.008)$.
The leading correlation would seem clearly significant and, indeed,
from our approximate formula \eqref{eq:thetatwapp}, $\theta_{0.99}^{\mathrm{TW}}
= 0.184$.

To assess the second correlation $r_2$, we appeal to the conservative
test discussed above based on the null distribution with $q-1 = 4, p
=4$ and $n = 138$. The Tracy--Widom approximation $\theta_{0.99}^{\mathrm{TW}}
\approx \mu + 2.023 \sigma \doteq 0.166 \ll 0.554$, which strongly
suggests that this second correlation is significant as well.

Marginal histograms naturally reveal some departures from symmetric
Gaussian tails, but they do not seem extreme enough to invalidate the
conclusions, which are also confirmed by permutation
tests.

\section{Tests of common means or variances}\label{sec:tests-common-means}

\subsection{Equality of means for common covariance}\label{sec:equalmeans}

Suppose that we have $k$ populations with independent data matrices
$\mathbf{X}_i$ consisting of $n_i$ observations drawn from an $N_p(
\bolds{\mu}_i, \bolds{\Sigma}_i)$ and put $n = \sum n_i$.
This is the one-way
multivariate analysis of variance illustrated in Example 1.1.
For testing the null hypothesis of equality of means
$H_0\dvtx  \bolds{\mu}_1 = \cdots = \bolds{\mu}_k$, we form, for
each population, the sample mean $\bar{\mathbf{x}}_i$ and covariance
matrix $\mathbf{S}_i$,
normalized so that $n_i \mathbf{S}_i \sim W_p(\bolds{\Sigma}, n_i-1)$.
The basic quantities are the within groups sum of squares $\mathbf{W}
= \sum n_i \mathbf{S}_i \sim W_p( \bolds{\Sigma}, n- k)$ and the
between group sum of squares $\mathbf{B} =
\sum n_i (\bar{\mathbf{x}}_i - \bar{\mathbf{x}})(\bar{\mathbf{x}}_i -
\bar{\mathbf{x}})' \sim W_p(\bolds{\Sigma}, k-1)$ under $H_0$,
independently of $ \mathbf{W}$.
The union-intersection test of $H_0$ uses the largest root of
$\mathbf{W}^{-1} \mathbf{B}$ or, equivalently, that of
$(\mathbf{W} + \mathbf{B})^{-1} \mathbf{B}$, and the latter has, under
$H_0$, the $\theta(p, n-k, k-1)$ distribution.

\subsection{Equality of covariance matrices}\label{sec:equal-covar-matr}

Suppose that independent samples from two normal distributions
$N_p(\bolds{\mu}_1, \bolds{\Sigma}_1)$ and
$N_p(\bolds{\mu}_2, \bolds{\Sigma}_2)$ lead to covariance
estimates $\mathbf{S}_i$ which are independent and Wishart
distributed on $n_i$ degrees of freedom: $n_i \mathbf{S}_i \sim
W_p(n_i,\bolds{\Sigma}_i)$ for $i = 1,
2$. Then the largest root test of the null hypothesis $H_0\dvtx
\bolds{\Sigma}_1 = \bolds{\Sigma}_2$ is based on the largest
eigenvalue $\theta$ of $(n_1 \mathbf{S}_1 + n_2
\mathbf{S}_2)^{-1} n_2 \mathbf{S}_2$, which under $H_0$ has
the $\theta(p, n_1, n_2)$ distribution \citeauthor{muir82} (\citeyear{muir82}), page~332.

\section{Multivariate linear model}\label{sec:mlm}

The multivariate linear model blends ideas well known from the
univariate setting with new elements introduced  by correlated multiple
responses.
In view of the breadth of models covered, and the variety of notation
in the literature and in the software, we review the setting in a little
more detail, beginning with the familiar model for a single response
\[
\mathbf{y} = \mathbf{X} \bolds{\beta} + \mathbf{u}.
\]
Here $\mathbf{y}$ is an $n \times 1$ column vector of observations on
a response variable, $\mathbf{X}$ is an $n \times q$ model matrix, and
$\mathbf{u}$ is an $n \times 1$ column vector of errors, assumed here
to be independent and identically distributed as $N(0,\sigma^2)$.
The $q \times 1$ vector $\bolds{\beta}$ of unknown parameters has the
least squares estimate---when $\mathbf{X}$ has full rank---given by
\[
\hat{ \bolds{\beta}} =  ( \mathbf{X}' \mathbf{X} )^{-1}
\mathbf{X}' \mathbf{y}.
\]

The error sum of squares $SS_E = ( \mathbf{y - X
\hat{\bolds{\beta}}})' ( \mathbf{y - X
\hat{\bolds{\beta}}}) = \mathbf{y' P y}$, where $\mathbf{P}$
denotes orthogonal projection onto the subspace orthogonal to the
columns of $\mathbf{X}$, it has rank $n-q$, and so $SS_E \sim
\chi^2_{(n-q)}$.

Consider the linear hypothesis $H_0\dvtx  \mathbf{C}_1 \bolds{\beta} =
0$, where $\mathbf{C}_1$ is a $g \times q$ matrix of rank $g$.
In the simplest example, $\mathbf{C}_1 = [ \mathbf{I}_g \ \mathbf{0}]$
extracts the first $g$ elements of $\bolds{\beta}$; more
generally, the rows of $\mathbf{C}_1$ are often contrasts among the
components of $\bolds{\beta}$.
To describe the standard $F$-test of $H_0$, let $\mathbf{C}_2$ be any
$(q-g) \times q$ matrix such that $\mathbf{C} =
{\mathbf{C}_1 \choose
\mathbf{C}_2}$
becomes an invertible $q \times q$ matrix.
We may then write
\[
\mathbf{X} \bolds{\beta} =
\bigl[ \matrix{\mathbf{X C}^{(1)} & \mathbf{X C}^{(2)}} \bigr]
\pmatrix{
\mathbf{C}_1 \bolds{\beta}  \cr
\mathbf{C}_2 \bolds{\beta}
},
\]
where we have partitioned $\mathbf{C}^{-1} = [ \mathbf{C}^{(1)} \ \mathbf{C}^{(2)} ]$ into blocks with $g$ and $q-g$
columns respectively.

Let $\mathbf{P}_1$ denote the orthogonal projection onto the subspace
orthogonal to the columns of $\mathbf{X C}^{(2)}$. We have the sum of
squares decomposition
\[
\mathbf{y' P}_1 \mathbf{y} = \mathbf{y' P y} + \mathbf{y'} (\mathbf{P}_1 - \mathbf{P}) \mathbf{y}
\]
and the hypothesis sum of squares for testing $H_0 : \mathbf{C}_1
\bolds{\beta} = 0$ is given by $SS_H = \mathbf{y' P}_2 \mathbf{y}$, with
$\mathbf{P}_2 = \mathbf{P}_1 - \mathbf{P}$.
The projection $\mathbf{P}_2$ has rank $g$ and so under $H_0$, $SS_H
\sim \chi^2_{(g)}$.
The projections $\mathbf{P}$ and $\mathbf{P}_2$ are orthogonal and so
the sums of squares have independent chi-squared distributions, and
under $H_0$ the traditional $F$-statistic
\[
F = \frac{ SS_H/g}{ SS_E/(n-q)} \sim F_{g,n-q}.
\]
Explicit expressions for the sums of squares are given by
\begin{eqnarray*}
SS_E & = & \mathbf{y' \bigl(I - X (X' X)^{-1} X' \bigr) y},  \\
SS_H & = & (\mathbf{C}_1 \hat{\bolds{\beta}})'
[\mathbf{C}_1 \mathbf{(X' X)^{-1}} \mathbf{C}_1'
]^{-1}  \mathbf{C}_1 \hat{\bolds{\beta}}.
\end{eqnarray*}

In the multivariate linear model,
\[
\mathbf{Y} = \mathbf{X} \mathbf{B} + \mathbf{U},
\]
the single response $\mathbf{y}$ is replaced by $p$ response vectors,
organized as columns of the $n \times p$ matrix $\mathbf{Y}$. The
model (or design) matrix $\mathbf{X}$ remains the same for each
response; however, there are separate vectors of unknown coefficients
and errors for each response; these are organized into a $q \times p$
matrix $\mathbf{B}$ of regression coefficients and an $n \times p$
matrix $\mathbf{E}$ of errors.
The multivariate aspect of the model is the assumption that the rows
of $\mathbf{U}$ are indepedent, with multivariate normal distribution
having mean $\mathbf{0}$ and common covariance matrix
$\bolds{\Sigma}$.
Thus, $\mathbf{U}$ is a normal data matrix of $n$ samples from
$N_p(\mathbf{0}, \bolds{\Sigma})$.
Assuming for now that the model matrix $\mathbf{X}$ has full rank, the
least squares estimator
\[
\hat{\mathbf{B}} = \mathbf{(X'X)^{-1} X'Y}.
\]

The linear hypothesis becomes
\[
H_0 \dvtx    \mathbf{C}_1 \mathbf{B} = \mathbf{0}.
\]
The sums of squares of the univariate case are replaced by hypothesis
and error sums of squares and products matrices:
\begin{eqnarray}\label{eq:EHdef}
E & = & \mathbf{Y' P Y} = \mathbf{Y' \bigl(I - X (X' X)^{-1} X' \bigr) Y}, \nonumber\\[-8pt]\\[-8pt]
H & = & \mathbf{Y' P}_2 \mathbf{Y}
= (\mathbf{C}_1 \hat{\mathbf{B}})'
[\mathbf{C}_1 \mathbf{(X' X)^{-1}} \mathbf{C}_1']^{-1}
\mathbf{C}_1 \hat{\mathbf{B}},\nonumber
\end{eqnarray}
in which the univariate vectors $\mathbf{y}$ and
$\hat{\bolds{\beta}}$ are simply replaced by their multivariate
analogs $\mathbf{Y}$ and $\hat{\mathbf{B}}$.
It follows that $\mathbf{E} \sim W_p( \bolds{\Sigma}, n-q)$ and that
under $H_0$, $\mathbf{H} \sim W_p( \bolds{\Sigma}, g)$; furthermore,
$\mathbf{E}$ and $\mathbf{H}$ are independent.
Generalizations of the $F$-test are obtained from the eigenvalues
$(\lambda_i)$ of the matrix $\mathbf{E}^{-1} \mathbf{H}$ or,
equivalently, the eigenvalues $\theta_i$ of $(\mathbf{H+E})^{-1}
\mathbf{H}$.

Thus, under the null hypothesis $\mathbf{C}_1 \mathbf{B} =0$, Roy's
maximum root statistic $\theta_1$ has null distribution
\begin{eqnarray}\label{eq:pardef}
&&\theta_1 \sim \theta(p, n-q, g),  \qquad \mbox{where}\nonumber
\\
&&\qquad p  =  \mbox{dimension},\qquad  g  =  \operatorname{rank} (\mathbf{C}_1), \\
&&\hspace*{1pt}\qquad q  =  \operatorname{rank} (\mathbf{X}),\hspace*{10pt}\qquad  n  = \mbox{sample size}.
\nonumber
\end{eqnarray}


\subsection*{Two extensions}
(a) $X$ \textit{not of full rank.}
This situation routinely occurs when redundant parameterizations are
used, for example, when dealing with factors in analysis of variance
models. One approach (e.g., MKB, Section~6.4) is to rearrange the columns
of $\mathbf{X}$ and partition $\mathbf{X} = [\mathbf{X}_1 \ \mathbf{X}_2]$ so that $\mathbf{X}_1$ has full rank. We must also
assume that the matrix $\mathbf{C}_1$ is \textit{testable} in the
sense that, as a function of $\mathbf{B}$, $\mathbf{X B} = \mathbf{0}$
implies $\mathbf{C}_1 \mathbf{B} = \mathbf{0}$. In such cases, if we
partition $\mathbf{C}_1 = [ \mathbf{C}_{11} \ \mathbf{C}_{12} ]$
conformally with $\mathbf{X}$, then $\mathbf{C}_{12} = \mathbf{C}_{11}
(\mathbf{X}'_1 \mathbf{X}_1)^{-1} \mathbf{X}'_1 \mathbf{X}_2$ is
determined from $\mathbf{C}_{11}$.

With these assumptions, we use $\mathbf{X}_1$ and $\mathbf{C}_{11}$ in
(\ref{eq:EHdef}) and (\ref{eq:pardef}) in place of $\mathbf{X}$ and
$\mathbf{C}_1$.

(b) \textit{Intra-subject hypotheses.}
A straightforward extension is possible in order to test null
hypotheses of the form
\[
\mathbf{C}_1 \mathbf{B} \mathbf{M}_1 = 0,
\]
where $\textbf{M}_1$ is $p \times r$ of rank $r$. The columns of
$\mathbf{M}_1$ capture particular linear combinations of the dependent
variables---for an example, see, e.g., \citeauthor{morr05} (\citeyear{morr05}), Chapter~3.6.

We simply consider a modified linear model
\[
\mathbf{Y M}_1 = \mathbf{X B M}_1 + \mathbf{U M}_1.
\]
An important point is that the rows of $\mathbf{U M}_1$ are still
independent, now distributed as $N_p( \mathbf{0}, \mathbf{M}'_1
\bolds{\Sigma} \mathbf{M})$.
So we may simply apply the above analysis, replacing $\mathbf{Y},
\mathbf{U}$ and $\mathbf{B}$ by $\mathbf{Y M}_1$, $\mathbf{U M}_1$ and
$\mathbf{B M}_1$ respectively.
In particular, the greatest root statistic now has null distribution
given by
\[
\theta_1 \sim \theta( r, n-q, g).
\]

\subsection*{Linear hypotheses in SAS}
Analyses involving the four multivariate tests are provided in a
number of SAS routines, such as GLM and CANCORR.
The parameterization used here can be translated into that used in SAS
by means of the documentation given in the SAS/STAT Users Guide---we
refer to the section on Multivariate Tests in version 9.1, page 48 ff.
The linear hypotheses correspond to MKB notation via
\begin{center}
\vspace*{5pt}
\begin{tabular}{cc}
MKB & SAS \\
\hline
$\mathbf{C}_1$ & $\mathbf{L}$ \\
$\mathbf{B}$  &  $\bolds{\beta}$ \\
$\mathbf{M}_1$ & $\mathbf{M}$
\end{tabular}\hspace*{2pt}
\vspace*{5pt}
\end{center}
while the parameters of the greatest root distribution are given by
\begin{center}
\vspace*{5pt}
\begin{tabular}{l|ll|ll|}
& \multicolumn{2}{c|}{MKB} & \multicolumn{2}{c|}{SAS} \\
\hline
dimension & $r$ & $\operatorname{rank} (\textbf{M}_1) $ &
$\operatorname{rank} (\mathbf{M}) $ & $\mathsf{p}$   \\
hypothesis & $g$ &  $ \operatorname{rank}(\mathbf{C}_1)$   &
$\operatorname{rank} (\mathbf{L}) $ & $\mathsf{q}$   \\
error & $n-q$ & & & $\mathsf{v}$
\end{tabular}\hspace*{2pt}.
\vspace*{5pt}
\end{center}
(Note: we use \textsf{sans serif} font for the SAS parameters!)
Finally, the SAS printouts use the following parameters:
\begin{eqnarray*}
\mathsf{s} & = & \mathsf{p} \wedge \mathsf{q}, \\
\mathsf{m} & = & ( |\mathsf{p} - \mathsf{q}|-1)/2, \\
\mathsf{n} & = & ( \mathsf{v} - \mathsf{p} -1)/2.
\end{eqnarray*}

\section{Concluding discussion}\label{sec:concl-disc}

We have described the Tracy--Widom approximation to the null
distribution for the largest root test for a variety of classical
multivariate procedures. These procedures exhibit varying degrees of
sensitivity to the assumption of normality, independence etc.
Documenting the sensitivity/robustness of the T--W approximation is
clearly an important issue for further work.  Two brief remarks can be
made. In the corresponding single Wishart setting
[e.g., \citet{john00c}], the largest eigenvalue can be shown, under the null
distribution, to still have the T--W limit if the original data have
``light tails'' (i.e., sub-Gaussian) [see \citet{sosh01a};
\citet{pech07}].
In  the double Wishart settings, simulations for canonical correlation
analysis with $n=100$ samples on $q=20$ and $p=10$ variables, each
following i.i.d. $t_{(5)}$ or i.i.d. random sign distributions, showed
that the T--W distribution for the leading correlation $r_1^2$ still
holds in the central 99\% of the distribution.

\begin{appendix}
\section*{Appendix: Proof of lemma}\label{sec:concl-disc-1}

If $\bolds{\Sigma} \in H_s$, there are at most $s$ nonzero canonical
correlations, and we may suppose without loss of generality that the
$q$ $\mathbf{x}$-variables have been transformed so that only the last $s$ of
them have any correlation with $\mathbf{Y}$.
We employ the singular value decomposition (SVD) description of CCA,
cf. \citeauthor{govl96} (\citeyear{govl96}), Section~12.4.3.
Using QR decompositions, write
\[
\mathbf{X} = Q_X R_X,\qquad   \mathbf{Y} = Q_Y R_Y.
\]
Let $C = Q_X^T Q_Y$ and form the SVD $C = U R V^T$.
Then the diagonal elements $r_1 \geq r_2 \geq \cdots \geq r_{\min(p,q)}$ of
$R$ contain the sample canonical correlations.

Now consider the reduced $n \times (q-s)$ matrix $\mathbf{X}^-$
obtained by dropping the last $s$ columns from $\mathbf{X}$.
Form the QR decomposition $\mathbf{X}^- = Q_{X^-} R_{X^-}$. From the
nature of the decomposition, we have $Q_X = [ Q_{X^-} \ Q^+]$,
that is, $Q_{X^-}$ represents\vspace*{1pt} the first $q-s$ columns of $Q_X$.
Consequently, $C_- = Q_{X^-}^T Q_Y$ forms the first $q-s$ rows of $C$.
Our lemma now follows from the interlacing property of singular
values [e.g., \citeauthor{govl96} (\citeyear{govl96}),
Corollary~8.6.3].
\[
\sigma_{s+1}(C) \leq \sigma_1(C_-).
\]
Indeed, our earlier discussion implies that $\mathbf{X}^-$ and
$\mathbf{Y}$ are independent, and so $\sigma_1(C_-)$ has the null
distribution $\mathcal{L}(r_1 | p, q-s, n; \mathbf{I})$.
\end{appendix}
\section*{Acknowledgment}
William Chen graciously provided an electronic copy of his tables for the distribution of the largest root.

\printaddresses


\begin{thebibliography}{99}

\bibitem[\protect\citeauthoryear{Anderson}{2003}]{ande03}
\textsc{Anderson, T.~W.} (2003).
\textit{An Introduction to Multivariate Statistical Analysis}, 3rd ed. Wiley, Hoboken, NJ.
\MR{1990662}

\bibitem[\protect\citeauthoryear{Andrews and Herzberg}{1985}]{anhe85}
\textsc{Andrews, D.~F.} and \textsc{Herzberg, A.~M.} (1985).
\textit{Data}. Springer, New York.

\bibitem[\protect\citeauthoryear{Chen}{2002}]{chen02a}
\textsc{Chen, W.~W.} (2002).
Some new tables of the largest root of a matrix in multivariate analysis: A~computer approach from 2 to 6.
Presented at the 2002 American Statistical Association.

\bibitem[\protect\citeauthoryear{Chen}{2003}]{chen03}
\textsc{Chen, W.~W.} (2003).
Table for upper percentage points of the largest root of a determinantal equation with five roots.
\textit{InterStat} (5).
Available at \href{http://interstat.statjournals.net}{interstat.statjournals.net}.

\bibitem[\protect\citeauthoryear{Chen}{2004a}]{chen04}
\textsc{Chen, W.~W.} (2004a).
The new table for upper percentage points of the largest root of a determinantal equation with seven roots.
\textit{InterStat} (1).
Available at \href{http://interstat.statjournals.net}{interstat.statjournals.net}.

\bibitem[\protect\citeauthoryear{Chen}{2004b}]{chen04a}
\textsc{Chen, W.~W.} (2004b).
Some new tables for the upper probability points of the largest root of a determinantal
equation with seven and eight roots.
In \textit{Special Studies in Federal Tax Statistics. Statistics of Income Division, Internal Revenue Service}
(J.~Dalton and B.~Kilss, eds.) 113--116.


\bibitem[\protect\citeauthoryear{Constantine}{1963}]{cons63}
\textsc{Constantine, A.~G.} (1963).
Some non-central distribution problems in multivariate analysis.
\textit{Ann. Math. Statist.} \textbf{34}~1270--1285.
\MR{0181056}

\bibitem[\protect\citeauthoryear{Davis}{1972}]{davi72}
\textsc{Davis, A.~W.} (1972).
On the marginal distributions of the latent roots of the multivariate beta matrix.
\textit{Ann. Math. Statist.} \textbf{43}~1664--1670.
\MR{0343465}

\bibitem[\protect\citeauthoryear{Foster}{1957}]{fost57}
\textsc{Foster, F.~G.} (1957).
Upper percentage points of the generalized {B}eta distribution. {II}.
\textit{Biometrika} \textbf{44}~441--453.
\MR{0090199}

\bibitem[\protect\citeauthoryear{Foster}{1958}]{fost58}
\textsc{Foster, F.~G.} (1958).
Upper percentage
points of the generalized {B}eta distribution. {III}.
\textit{Biometrika} \textbf{ 45}~492--503.
\MR{0100366}

\bibitem[\protect\citeauthoryear{Foster and Rees}{1957}]{fore57}
\textsc{Foster, F.~G.} and \textsc{Rees, D.~H.} (1957).
Upper percentage points of the generalized {B}eta distribution. {I}.
\textit{Biometrika} \textbf{44}~237--247.
\MR{0086462}

\bibitem[\protect\citeauthoryear{Golub and Van~{L}oan}{1996}]{govl96}
\textsc{Golub, G.~H.} and \textsc{Van~{L}oan, C.~F.} (1996).
\textit{Matrix Computations}, 3rd ed.
Johns Hopkins Univ. Press, Baltimore.
\MR{1417720}

\bibitem[\protect\citeauthoryear{Heck}{1960}]{heck60}
\textsc{Heck, D.~L.} (1960).
Charts of some upper
percentage points of the distribution of the largest characteristic root.
\textit{Ann. Math. Statist.} \textbf{31}~625--642.
\MR{0119301}

\bibitem[\protect\citeauthoryear{Johnson and Wichern}{2002}]{jowi07}
\textsc{Johnson, R.~A.} and \textsc{Wichern, D.~W.} (2002).
\textit{Applied Multivariate Statistical Analysis}, 6th ed. Pearson Prentice Hall, Upper Saddle River, NJ.

\bibitem[\protect\citeauthoryear{Johnstone}{2001}]{john00c}
\textsc{Johnstone, I.~M.} (2001).
On the distribution
of the largest eigenvalue in principal components analysis.
\textit{Ann. Statist.} \textbf{29}~295--327.
\MR{1863961}

\bibitem[\protect\citeauthoryear{Johnstone}{2008}]{john08}
\textsc{Johnstone, I.~M.} (2008).
Multivariate
analysis and {J}acobi ensembles: Largest eigenvalue, Tracy--{W}idom limits and
rates of convergence.
\textit{Ann. Statist.}
\textbf{36} 2638--2716.
\MR{2485010}

\bibitem[\protect\citeauthoryear{Johnstone and Chen}{2007}]{joch07}
\textsc{Johnstone, I.~M.} and \textsc{Chen, W.~W.} (2007).
Finite sample accuracy of {T}racy--{W}idom
approximation for multivariate analysis.
In \textit{2007 JSM Proceedings} 1161--1166.
Amer. Statist. Assoc., Alexandria, VA.

\bibitem[\protect\citeauthoryear{Johnstone et~al.}{2010}]{jmps10}
\textsc{Johnstone, I.~M., Ma, Z., Perry, P.~O.} and \textsc{Shahram, M.} (2010).
\texttt{RMTstat}: Distributions,
statistics and tests derived from random matrix theory.
Manuscript in preparation.

\bibitem[\protect\citeauthoryear{Koev}{2010}]{koev07}
\textsc{Koev, P.} (2010).
Computing multivariate statistics.
Manuscript in preparation.

\bibitem[\protect\citeauthoryear{Koev and Edelman}{2006}]{koed06}
\textsc{Koev, P.} and \textsc{Edelman, A.} (2006).
The efficient evaluation of the hypergeometric function of a matrix argument.
\textit{Math. Comp.} \textbf{75}~833--846 (electronic).
\MR{2196994}

\bibitem[\protect\citeauthoryear{Krishnaiah}{1980}]{kris80}
\textsc{Krishnaiah, P.~R.} (1980).
Computations of some multivariate distributions.
In \textit{Handbook of Statistics, Volume 1---Analysis of Variance}
(P.~R. Krishnaiah, ed.) 745--971.
North-Holland, Amsterdam.

\bibitem[\protect\citeauthoryear{Lutz}{1992}]{lutz92}
\textsc{Lutz, J.~G.} (1992).
A {T}urbo {P}ascal unit
for approximating the cumulative distribution function of {R}oy's largest
root criterion.
\textit{Educational and Psychological Measurement} \textbf{52}~899--904.

\bibitem[\protect\citeauthoryear{Lutz}{2000}]{lutz00}
\textsc{Lutz, J.~G.} (2000).
Roy table: A program
for generating tables of critical values for {R}oy's largest root criterion.
\textit{Educational and Psychological Measurement} \textbf{60}~644--647.

\bibitem[\protect\citeauthoryear{Mardia, Kent and Bibby}{1979}]{mkb79}
\textsc{Mardia, K.~V., Kent, J.~T.} and \textsc{Bibby, J.~M.} (1979).
\textit{Multivariate Analysis}.
Academic Press, London.
\MR{0560319}

\bibitem[\protect\citeauthoryear{Morrison}{2005}]{morr05}
\textsc{Morrison, D.~F.} (2005).
\textit{Multivariate Statistical Methods}, 4th ed. Thomson, Belmont, CA.

\bibitem[\protect\citeauthoryear{Muirhead}{1982}]{muir82}
\textsc{Muirhead, R.~J.} (1982).
\textit{Aspects of Multivariate Statistical Theory}. Wiley, New York.
\MR{0652932}

\bibitem[\protect\citeauthoryear{Nanda}{1948}]{nand48}
\textsc{Nanda, D.~N.} (1948).
Distribution of a root of a determinantal equation.
\textit{Ann. Math. Statist.} \textbf{19}~47--57.
\MR{0024106}

\bibitem[\protect\citeauthoryear{Nanda}{1951}]{nand51}
\textsc{Nanda, D.~N.} (1951).
Probability distribution tables of the largest root of a determinantal equation with two roots.
\textit{J.~Indian Soc. Agricultural Statist.} \textbf{3}~175--177.
\MR{0044788}

\bibitem[\protect\citeauthoryear{P\'{e}ch\'{e}}{2009}]{pech07}
\textsc{P\'{e}ch\'{e}, S.} (2009).
Universality results for largest eigenvalues of some sample covariance matrices ensembles.
\textit{Probab. Theory Related Fields} \textbf{143} 481--516. 

\bibitem[\protect\citeauthoryear{Pillai}{1955}]{pill55}
\textsc{Pillai, K. C.~S.} (1955).
Some new test criteria in multivariate analysis.
\textit{Ann. Math. Statist.} \textbf{26}~117--121.
\MR{0067429}

\bibitem[\protect\citeauthoryear{Pillai}{1956a}]{pill56}
\textsc{Pillai, K. C.~S.} (1956a).
On the distribution of the largest or smallest root of a matrix in multivariate analysis.
\textit{Biometrika} \textbf{43}~122--127.
\MR{0077052}

\bibitem[\protect\citeauthoryear{Pillai}{1956b}]{pill56a}
\textsc{Pillai, K. C.~S.} (1956b).
Some results useful in multivariate analysis.
\textit{Ann. Math. Statist.} \textbf{27}~1106--1114.
\MR{0081852}

\bibitem[\protect\citeauthoryear{Pillai}{1957}]{pill57}
\textsc{Pillai, K. C.~S.} (1957).
\textit{Concise Tables for Statisticians}.
The Statistical Center, Univ. of the Philippines,
Manila.

\bibitem[\protect\citeauthoryear{Pillai}{1965}]{pill65}
\textsc{Pillai, K. C.~S.} (1965).
On the distribution of the largest characteristic root of a matrix in multivariate analysis.
\textit{Biometrika} \textbf{52}~405--414.
\MR{0205391}

\bibitem[\protect\citeauthoryear{Pillai}{1967}]{pill67}
\textsc{Pillai, K. C.~S.} (1967).
Upper percentage points of the largest root of a matrix in multivariate analysis.
\textit{Biometrika} \textbf{54}~189--194.
\MR{0215433}

\bibitem[\protect\citeauthoryear{Pillai and Bantegui}{1959}]{pill59}
\textsc{Pillai, K. C.~S.} and \textsc{Bantegui, C.~G.} (1959).
On the distribution of the largest of six roots of a
matrix in multivariate analysis.
\textit{Biometrika} \textbf{46}~237--240.
\MR{0102152}

\bibitem[\protect\citeauthoryear{Pillai and Flury}{1984}]{pifl84}
\textsc{Pillai, K. C.~S.} and \textsc{Flury, B.~N.} (1984).
Percentage points of the largest characteristic root
of the multivariate beta matrix.
\textit{Commun. Statist. Part A} \textbf{13}~2199--2237.
\MR{0754832}

\bibitem[\protect\citeauthoryear{Rencher}{2002}]{renc02}
\textsc{Rencher, A.~C.} (2002).
\textit{Methods of  Multivariate Analysis}, 2nd ed.
Wiley, New York.
\MR{1885894}

\bibitem[\protect\citeauthoryear{Soshnikov}{2002}]{sosh01a}
\textsc{Soshnikov, A.} (2002).
A note on universality of the distribution of the largest eigenvalues in certain classes of sample
covariance matrices.
\textit{J.~Statist. Phys.} \textbf{108}~1033--1056.
\MR{1933444}

\bibitem[\protect\citeauthoryear{Tracy and Widom}{1996}]{trwi96}
\textsc{Tracy, C.~A.} and \textsc{Widom, H.} (1996).
On orthogonal and symplectic matrix ensembles.
\textit{Commun. Math. Phys.} \textbf{177}~727--754.
\MR{1385083}

\bibitem[\protect\citeauthoryear{Waugh}{1942}]{waug42}
\textsc{Waugh, F.~V.} (1942).
Regressions between sets of variables.
\textit{Econometrica} \textbf{10}~290--310.
\end{thebibliography}
\end{document}